# Necklace Beam Generation in Nonlinear Colloidal Engineered Media

SALIH Z. SILAHLI,[1] WIKTOR WALASIK,[1] NATALIA M. LITCHINITSER[1,*]

[1]Electrical Engineering Department, University at Buffalo, The State University of New York, Buffalo, NY 14260
*Corresponding author: natashal@buffalo.edu

**Modulational instability is a phenomenon that reveals itself as the exponential growth of weak perturbations in the presence of an intense pump beam propagating in a nonlinear medium. It plays a key role in such nonlinear optical processes as supercontinuum generation, light filamentation, and rogue waves. However, practical realization of these phenomena in the majority of available nonlinear media still relies on high-intensity optical beams. Here, we analytically and numerically show the possibility of necklace beam generation originating from low-intensity spatial modulational instability of vortex beams in engineered soft-matter nonlinear media.**

© 2015 Optical Society of America

http://dx.doi.org/10.1364/OL.99.099999

The emergence of metamaterials offers a strong potential to enable a plethora of novel nonlinear light-matter interactions and even new nonlinear materials. In particular, nonlinear focusing and defocusing effects are of paramount importance for manipulation of the minimum focusing spot size of structured light beams necessary for complex beam shaping, manipulation, and light filamentation.

While the quest for highly nonlinear materials for low-intensity nonlinear optics has been underway for decades, nonlinear optical materials available to date are still limited by relatively low and generally band-limited nonlinear susceptibilities responsible for ultrafast nonlinear processes. Many recent studies of nonlinear optics focused on the plasmonic as well as other high-Q resonantly enhanced structures for nonlinear optics that certainly have a strong potential for such applications as sensing or spectroscopy.

However, a number of nonlinear optical applications, including self-guiding, all-optical switching or wavelength conversion requiring high efficiency, still call for materials with significantly larger and tunable nonlinear response. The emergence of metamaterials has a potential to provide a breakthrough in the development of such materials.

It has already been demonstrated that linear optical properties of metamaterials, such as dielectric permittivity, magnetic permeability, and refractive index of metamaterials, can be designed to be positive, negative, or even zero at any selected frequency. This tenability extends also to a nonlinear regime, where nonlinear properties of metamaterials can be largely transformed by properly adjusting the dimensions, periodicity, and other parameters of the so-called meta-atoms (the unit cells of metamaterials) [1]. Indeed, metamaterials were predicted to fundamentally change fundamental nonlinear processes, including second harmonic generation, soliton propagation, four-wave mixing, modulation instability, and optical bistability, to name a few. In this study, we focus on modulational instability, a phenomenon that reveals itself as the exponential growth of weak perturbations when an intense pump beam propagates inside a nonlinear medium. It was shown to be closely related to such important phenomena as supercontinuum generation, light filamentation, and rogue waves [2-9].

To date, a majority of studies in the field of metamaterials have focused on solid-state nanostructures. However, engineering of optical properties in soft-matter offers a number of new degrees of freedom for designing optical polarizabilites. While the first ideas of using interplay of optical forces [10] and nonlinear self-action effects were proposed in 1980s [11], recently, significant progress and new insights have been made. Indeed, it has been proposed that colloidal suspensions offer a promising platform for engineering polarizabilities and realization of large and tunable nonlinearities [12-19]. In this paper, we analytically and numerically study the phenomenon of spatial modulational instability (MI) in soft matter nonlinear medium using beams with an orbital angular momentum (OAM) and predict the possibility of necklace beam generation originating at relatively low pump beam intensities.

In the majority of studies of nonlinear optical effects in colloidal suspensions, the optical nonlinearity was assumed to be of the Kerr type. However, recent studies of such system have shown that this rather simplified assumption is only valid when the optical beam intensity is kept below the threshold value determined by the thermal energy. In a more general case, the nonlinearity turns out to be exponential [13]. Moreover, this exponential nonlinearity can be saturable or super-critical with intensity, depending on the sign of the polarizability of the particles in the suspension.

In order to understand the reason behind these different types of nonlinearity, we note that colloidal suspensions can be created of two kinds of particles: those with positive polarizabilities (PP) corresponding to a particle refractive index greater than the background index ($n_p > n_b$) or those with negative polarizability (NP), with the corresponding refractive index of a particle lower than that of the background material ($n_p < n_b$) [4]. PP dielectric particles are attracted towards the high-intensity region of the beam, whereas in NP, particles are repelled. At the same time, in the PP case, the nonlinear scattering losses increase in the high-intensity region of the beam. In contrast, in the NP case, the nonlinear losses decrease with the increase of intensity since the particle concentration decreases.

While a majority of previous studies investigated nonlinear propagation of Gaussian beams in colloidal suspensions, here, we analyze the azimuthal MI for optical vortices with different topological charges propagating in the nano-colloidal system. We show that different types of the exponential optical nonlinearity in the PP and NP cases lead to different MI gain for the same perturbation and to the formation of different spatial distributions of the necklace beams.

Let us consider an optical beam propagating in the nano-colloidal system. The nonlinear Schrödinger equation governing the evolution of the slowly varying electric field envelope $\varphi$ can be written as [17,18]

$$i\frac{\partial \varphi}{\partial Z} + \frac{1}{2k_0 n_b}\nabla_\perp^2 \varphi + k_0(n_b - n_p)V_p\rho_0 e^{\frac{\alpha}{4k_B T}|\varphi|^2}\varphi + \frac{i}{2}\sigma\rho_0 e^{\frac{\alpha}{4k_B T}|\varphi|^2}\varphi = 0. \quad (1)$$

The particle polarizability is denoted by $\alpha$, and $k_B T$ is the thermal energy. $n_p$ and $n_b$ are particle and background refractive indices, respectively. $V_p$ is the volume of a particle, $\rho_0$ is the unperturbed particle concentration, $\sigma$ is the scattering cross-section, and $k_0 = 2\pi/\lambda_0$ is the wave number. After the normalization, Eq. (1) takes the following form:

$$i\frac{\partial U}{\partial \xi} + \nabla_\perp^2 U + (a + i\delta)\exp[a|U|^2]U = 0, \quad (2)$$

where $U$ is the normalized field amplitude, $\delta$ is normalized loss coefficient, and $\xi$ is the normalized propagation distance. For particles with PP, $n_p > n_b$ and the normalized nonlinear parameter $a=+1$. Consequently, the nonlinearity in the PP case is super-critical. For particles with NP, $n_p < n_b$ and the normalized nonlinear parameter is $a=-1$. For NP, the nonlinearity is saturable. The modulation instability is analyzed by applying a method involving a linear stability analysis and described in Ref. [3] to Eq. (2). The azimuthal phase dependency of the vortex beams can be easily expressed in the cylindrical coordinate frame. Therefore, it is convenient to express the transverse Laplacian in Eq. (2) in this frame:

$$i\frac{\partial U}{\partial \xi} + \left(\frac{1}{r}\frac{\partial U}{\partial r} + \frac{\partial^2 U}{\partial r^2} + \frac{1}{r^2}\frac{\partial^2 U}{\partial \theta^2}\right) + a\exp[a|U|^2]U = 0. \quad (3)$$

In Eq. (3), we neglect the loss term in order to facilitate the analytical studies of the MI. The normalized propagation distance is $\xi = Z/(2k_0^2 n_b L^2)$, the normalized radial coordinate $r = R/L$, and $L = (2k_0^2 n_b |n_p - n_b| V_p \rho_0)^{-1/2}$.

Next, we perform a standard linear stability analysis of the OAM beam propagating in a nano-colloidal system. We apply azimuthal perturbations to the steady state solution $U_{st}(r, \theta)$ of Eq. (3). Here, perturbations are applied only to the azimuthal field distribution taken at the radial distance for which the intensity is constant $U_0(\theta) = U_{st}(r = r_m, \theta)$, where $r_m$ defines a mean radius of the steady state solution $U_{st}(r, \theta)$ and is calculated using Eq. (21) in Ref. [3]. The perturbed field distribution is given by

$$U_p(\xi, \theta) = \left(|U_0| + a_1 e^{-i(M\theta + \mu\xi)} + a_2^* e^{i(M\theta + \mu^*\xi)}\right)e^{i\lambda\xi + im\theta}, \quad (4)$$

where $|U_0|$ is the amplitude of the steady state solution, $a_1, a_2$ are the amplitudes of the small perturbations ($a_1, a_2 \ll |U_0|$), and $m$ and $M$ are the azimuthal indices of the steady state solution (topological vortex charge) and the perturbation, respectively. $\lambda$ is the propagation constant of the steady state solution, and $\mu$ is the propagation constant correction for the perturbation.

Substituting Eq. (4) into Eq. (3) and linearizing the equation in $a_1$ and $a_2$, we obtain the following coupled equations for perturbation amplitudes:

$$-a_1(\lambda - \mu) - \frac{(m + M)^2}{r_m^2}a_1 + f(|U_0|^2)a_1 + f'(|U_0|^2)|U_0|^2(a_1 + a_2) = 0, \quad (5)$$

$$-a_2^*(\lambda + \mu^*) - \frac{(m - M)^2}{r_m^2}a_2^* + f(U_0^2)a_2^* + f'(|U_0|^2)|U_0|^2(a_1^* + a_2^*) = 0. \quad (6)$$

Here, $f(|U|^2) = a\exp(a|U|^2)$ and a prime denotes the derivative with respect to $|U|^2$. Equation (5) and the complex conjugate of Eq. (6) can be written as an eigenvalue problem in the following matrix form:

$$\begin{bmatrix} A + \mu & B \\ -B & C + \mu \end{bmatrix}\begin{bmatrix} a_1 \\ a_2 \end{bmatrix} = 0, \quad (7)$$

where

$A = -\lambda - \frac{(m+M)^2}{r_m^2} + f(|U_0|^2) + f'(|U_0|^2)|U_0|^2$;

$B = f'(|U_0|^2)|U_0|^2$; and

$C = \lambda + \frac{(m-M)^2}{r_m^2} - f(|U_0|^2) - f'(|U_0|^2)|U_0|^2$.

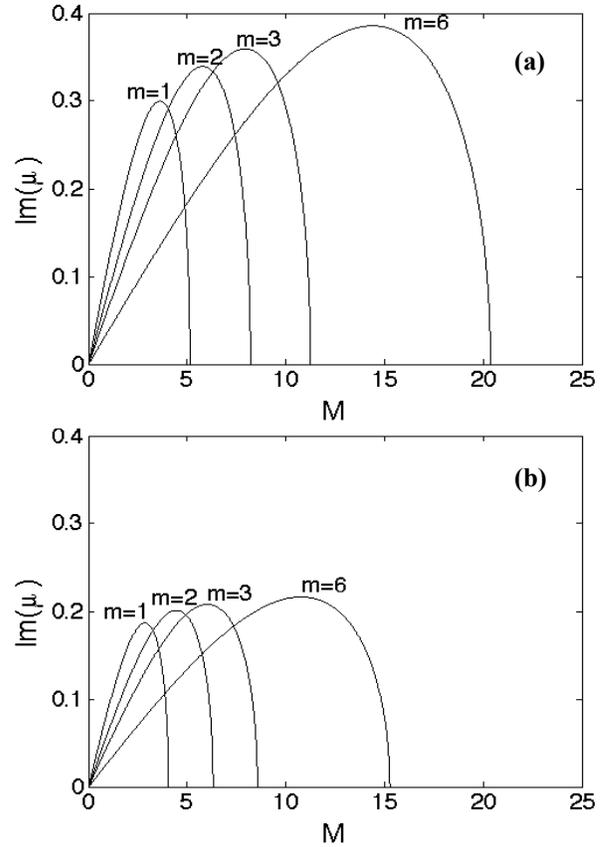

**Fig 1**. Azimuthal modulation instability gain $Im(\mu)$ as a function of the perturbation azimuthal index $M$, for (a) PP and (b) NP particle-based systems for different topological charges $m$ of the initial steady state vortex solution.

Equation (7) has nontrivial solutions when the matrix determinant is equal to zero. This condition yields two expressions: (i) for the propagation constant:

$$\lambda = -\frac{m^2}{r_m^2} + f(|U_0|^2), \quad (8)$$

and (ii) for the propagation constant correction $\mu$ associated with the MI. The MI gain is given by the imaginary part of $\mu$:

$$Im(\mu) = \frac{M}{r_m} \times Im \sqrt{\frac{M^2}{r_m^2} - 2|U_0|^2 \exp(a|U_0|^2)}. \quad (9)$$

The difference between the gain amplitude $Im(\mu)$ in the PP and NP case is the consequence of a different value of the parameter $a$ depending on the sign of polarizability.

The result presented as Eq. (9) is used in the following to predict the MI gain for vortices propagating in the nano-colloidal media. We study the propagation of light with the wavelength $\lambda_0 = 532$ nm, in two systems. The PP suspension composed of high refractive index ($n_p = 1.56$) polystyrene particles randomly dispersed in water (the background refractive index $n_b=1.33$) with the volume filling factor $(f = V_p \rho_0) f = 1.4 \cdot 10^{-3}$. The NP suspension made of low refractive index particles (air bubbles with $n_p=1$) dispersed in water with the volume filling fraction $f = 10^{-3}$. In both cases, the radius of the particles is assumed to be 50 nm. For these two sets of the parameters, Fig. 1 shows the gain curves $Im(\mu)$ as a function of the perturbation azimuthal index $M$ for different values of the vortex charge $m$.

We observe that for a fixed vortex charge $m$, the MI gain is higher in the PP particle-based system than in the NP particle-based system. As a result, for a fixed initial power, the MI onsets for a shorter propagation distance for PP than for the NP case. Additionally, we observe that the number of peaks in the necklace beam, associated with the perturbation azimuthal index $M$ for which $Im(\mu)$ is maximal, is larger for the PP case than for the NP case, for a fixed initial vortex charge.

The theoretical predictions based on the analytical expression for the MI gain [Eq. (9)] are verified by a direct numerical solution of Eq. (1). This equation is solved using a three-dimensional split-step Fourier algorithm [20,21] in order to study the vortex dynamics in nano-colloidal systems. The input filed in the numerical algorithm corresponds to the steady state solution of Eq. (1) $U_{st}(r,\theta)$ for the $m$-th order optical vortex written in physical units:

$$\varphi(r,\theta,z=0) = A_m (r/\omega_0)^m \exp^{-r^2/(2r_m^2)+im\theta}, \quad (10)$$

where the width $\omega_0$ for the $m$-th order stable vortex is related to the averaged radius $r_m$ as $\omega_0 = r_m/(m+1)$ and the amplitude $A_m$ can be deduced from the total power of the stable solution

$$P_m = \frac{4\pi k_B T}{|\alpha|} \frac{2^{2m+1} m! (m+1)!}{(2m)!} L^2 \varepsilon_0 c n_b, \quad (11)$$

where $c$ denotes the speed of light in vacuum and $\varepsilon_0$ is the vacuum permittivity. In order to accelerate the growth of the MI, 10% of random noise is added to the input field.

Figures 2 and 3 show the dynamics of light propagation of vortex beams with various topological charges $m$ for PP and NP particles, respectively. The first column presents the steady state solution with the noise, the second column shows the vortex at half of the distance corresponding to the MI onset, and the third column presents the generated necklace beams immediately after the onset of the MI breakup. There is good agreement between the analytical predictions shown in Fig. 1 and numerical simulations. Indeed, for a given charge $m$, the number of maxima observed in the necklace beam generated by the numerical simulations corresponds to the azimuthal frequency $M$ with the highest grow rate predicted analytically. Moreover, in excellent agreement with analytical predictions, numerical simulations show that the necklace beam formation occurs at shorter distances in the PP case than in the NP case. The MI gain is higher for the PP particle-based medium due to the different type of nonlinearity.

In both PP and NP cases, the distance at which the MI appears decreases with the increase of the vortex topological charge. The distance required to observe the MI for PP case is more than 3 times shorter than that for the NP case. However, the analytical predictions suggest that this distance should be only two times shorter. Another discrepancy between the analytical predictions and the numerical calculations is the power required to observe the steady state solution in the case of NP particles. These discrepancies can be explained based on the assumptions made in the analytical calculations.

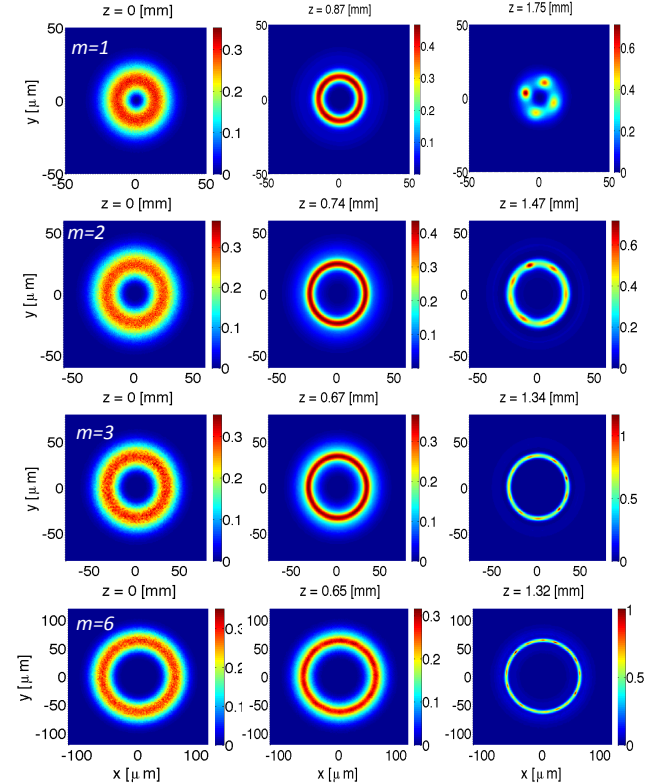

**Fig 2**. Distribution of the electric field intensity $|\varphi|^2$ [V$^2$/m$^2$] presenting the dynamics of the necklace beams for different topological charges $m$=1,2,3 and 6 in the PP particle-based system.

The values of the power corresponding to the steady state solutions in the NP system calculated using Eq. (11) are $P_1$=3.39 W, $P_2$=6.78 W, $P_3$=10.86 W, and $P_6$ =26 W. These values were obtained neglecting losses and taking only three first terms of the Taylor expansion of the nonlinear function $f(|U|^2)$. In the system with loss and a full exponential nonlinearity, the steady state solutions computed in the approximate analytical way are not fully stable. The width of these solutions oscillates during the propagation, and they experience loss. Therefore, in order to observe the MI in the numerical simulations, we need to modify the stable solutions found analytically. We found out that increasing the analytically predicted input power for each topological charge by multiplying it by the same factor of 2.4 while keeping all the other parameters unchanged, suffices to observe the beam breakup. The input power levels in the numerical simulations were increased to $P_1$=8.25 W, $P_2$=16.5 W, $P_3$ =26.4 W, and $P_6$ =63 W. The results of the simulations using these power levels are shown in Figs. 2 and 3.

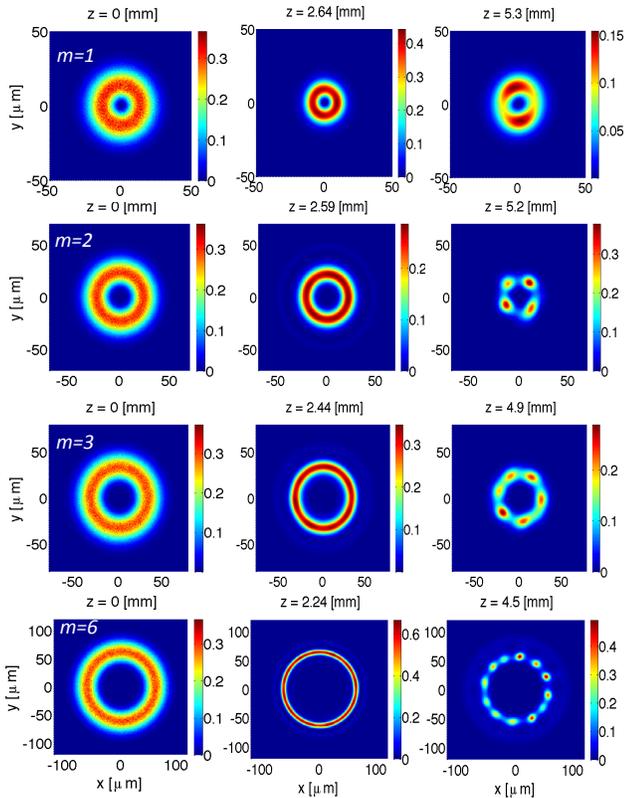

**Fig 3**. Distribution of the electric field intensity $|\varphi|^2$ [V$^2$/m$^2$] presenting the dynamics of the necklace beams for different topological charges $m=1,2,3$ and 6 in the NP particle-based system. The input vortices are the same as in the corresponding cases presented in Fig. 2.

In summary, we investigated the phenomenon of spatial modulational instability of OAM beams in colloidal nonlinear medium consisting of positive or negative polarizability particles and predict the formation of the necklace beams. We show that different types of exponential nonlinearity (saturable in the NP case and super-critical in the PP case) lead to different MI gains for the same perturbation. As a result, for the fixed input beam power, the MI onsets at a shorter propagation distance for PP as compared to the NP case. Also, the number of peaks in the necklace beam, associated with the perturbation azimuthal index *M*, is lager for the PP case than for the NP case, for a fixed initial vortex charge. These results may be of great importance to future studies of nonlinear propagation of structured light beams and filamentation in liquids. They can also be useful in applications of light propagation in highly scattering biological and chemical colloidal media.

This work was supported by the US Army Research office award W911NF-15-1-0146.